# FPGA Implementation of LS Code Generator for CDM Based MIMO Channel Sounder

M. Habib Ullah, Md. Niamul Bari, A. Unggul Priantoro

**Abstract**— MIMO (Multi Input Multi Output) wireless communication system is an innovative solution to improve the bandwidth efficiency by exploiting multipath-richness of the propagation environment. The degree of multipath-richness of the channel will determine the capacity gain attainable by MIMO deployment. Therefore, it is very important to have accurate knowledge of the propagation environment/radio channel before MIMO implement. The radio channel behavior can be estimated by channel measurement or channel sounding. CDM (Code Division multiplexing) is one of the channel sounding techniques that allow accurate measurement at the cost of hardware complexity. CDM based channel sounder, requires code with excellent auto-correlation and cross-correlation properties which generally difficult to achieve simultaneously. Theoretical analysis and computer simulation result demonstrated that, having excellent correlation propertied Loosely Synchronous (LS) code sequence perform efficiently. Finally, the an efficient LS code generator as a data source for transmitter implemented in Xilinx FPGA that can be integrated into CDM based 2x2 MIMO complete channel sounder.

---

## 1 INTRODUCTION

Multiple antenna array technologies can significantly enhance the performance of radio systems. MIMO signaling techniques in particular offer diversity/multiplexing gains to provide considerably higher channel throughputs as compared to conventional single antenna systems. For future wireless communication systems, MIMO techniques can be considered to provide high data throughput as well as significant enhancement in link reliability over existing systems. In real environments, however, accurate knowledge of the propagation channel is required to process MIMO receiving systems. Thus, the accuracy of MIMO channel measurement is an important issue in many aspects like simulation, system design, and performance analysis for beyond the third generation wireless communication systems

The previous schemes for MIMO channel measurement are to use time-division multiplexing with synchronized switching (TDMS) based MIMO channel sounder. Although this technique is cost effective, it has the major drawback that absolute time synchronization and excess time slots are needed, considering that each channel uses its own time slot [3]. In other words, the required time slots depend on the number of transmit antennas. As another approach, code-division multiplexing (CDM) based MIMO channel sounder with low correlation codes was introduced. However, it also has disadvantage that dynamic range is limited by the number of transmit antennas due to none-zero correlation values [5]. It has been proposed that using multiple transmit and receive antennas [20], and associated coding techniques could increase the performance of wireless communication systems. So far there has been a lot of theoretical research but relatively few practical systems have been demonstrated. Knowing the characteristic of the MIMO channel parameters is very important issue to determine performance of the MIMO system. The accurate knowledge of channel behavior is most important for efficient channel measurement.

In this paper a new efficient CDM-based MIMO channel sounding technique with LS code has been proposed [5]. Simulation results shows that the channel sounding scheme using LS codes gives very good performance for measuring $2 \times 2$ MIMO channel behavior. Fig.1 illustrates 2x2 MIMO system.

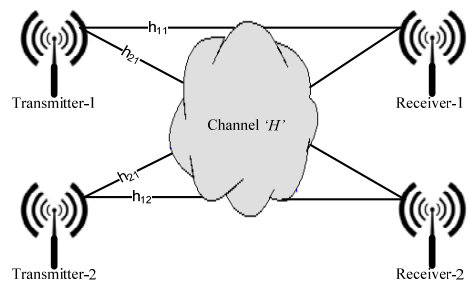

Fig.1. Illustration of 2×2 Multi-input Multi-output system.

## 2 OVERVIEW OF MIMO CHANNEL SOUDER

The characteristic of the MIMO channel is the vital issue to determine performance of the MIMO system. The accurate knowledge of channel behavior is most important for efficient mobile communication system.

The MIMO channel is considered to be parameterized by time-delay, complex path weight, direction of arrival (DOA), direction of departure (DOD), which can be obtained by analyzing measurement data. So, the accuracy of MIMO channel measurement is an important issue in many aspects like simulation, system design, and performance analysis. Channel sounders should therefore provide the

---

- M. Habib Ullah is student of Master in Communication Engineering Department of ECE, Faculty of Engineering, International Islamic University Malaysia.
- Md. Niamul Bari is sudent of Master in Computer and Information Engineering, Dept. of ECE, Faculty of Engineering, International Islamic University Malaysia
- A. Unggul Priantoro is assistant professor of Department of ECE, Faculty of Engineering, International Islamic University Malaysia.

temporal and spatial characteristics of the MIMO channel with accuracy and resolution high enough to befit the design purpose. In the conventional TDM based MIMO channel sounding technique [22, 23], the number of antennas at both transmitter and receiver limits its capability, with a trade-off between spatial resolution and time resolution. However, for the sake of reduced cost and complexity, most of the commercial MIMO channel sounders use single, time multiplexed, transceiver architecture [8].

The previous scheme for MIMO channel sounder is based on Time-Division Multiplexing(TDM) architecture that sounding signals from each transmit antenna are transmitted utilizing synchronized switch sequentially, not concurrently and uses PN sequence as a sounding signal. Although it has a merit of cost-effective in hardware implementation, but this scheme is not suitable for real-time channel measurement and has some major drawbacks such as the requirement of precise synchronization between transmitter and receiver and accuracy reduction during switching time as well. As a result, reformation about architecture of MIMO channel sounder is needed for overcoming these limitations of conventional TDM based MIMO channel sounder.

Another approach is CDM based MIMO channel sounding technique, which has the merit of real-time measurement [5]. In CDM architecture, sounding signals from all transmit antennas are transmitted simultaneously enables to measure the real-time MIMO channel parameters. The level of interference among different codes has a dominant effect on the performance of channel measurement [15]. However this scheme multiplexes transmit signals by using codes. Therefore it needs a code which has very low autocorrelation and cross-correlation values between different code sets for this CDM architecture.

## 3. PROPOSED DESIGN

In this thesis, CDM based Channel sounding technique proposed. As a solution of conventional TDM based channel sounder limitations, code division multiplexing (CDM) based MIMO channel sounding architecture is considered. One of the efficient codes is loosely synchronous (LS) codes with excellent correlation properties based on Golay complementary codes proposed.

A BPSK Transmitter with LS Code generator has been designed using the Sundance FPGA development boards.

The block diagram of the BPSK transmitter is shown in

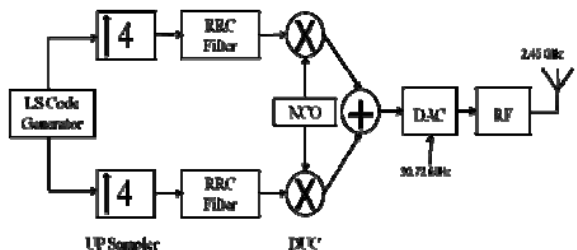

Fig.2 Block diagram of BPSK transmitter.

fig. 2. The transmitter is comprised of LS code sequence generator, up-sampler, Root Raised Cosine (RRC) filter, digital up-converter (DUC) and digital to analog converter (16 bit DAC) and RF transmitter. The RF is configured to transmit at the center frequency of 2.45 GHz. The LS code generator is used as the data source for the transmitter. The detail about LS code generator implementation is given in next section. The proposed design block diagram and explanation is given bellow:

### 3.1. LS Code Generator Design

The conventional schemes for MIMO channel measurement are to use time-division multiplexing with synchronized switching (TDMS) based MIMO channel sounder. Although this technique is cost effective, it has the major drawback that absolute time synchronization and excess time slots are needed, considering that each channel uses its own time slot. In other words, the required time slots depend on the number of transmit antennas. As another approach, code-division multiplexing (CDM) based MIMO channel sounder with low correlation codes was introduced. Due to the excellent correlation properties of the Loosely Synchronous Code, it can solve the problem caused in TDMA system. In CDM system it is really challenging to have such code sequence with good auto and cross correlation property at the same time. After simulation I have achieved the excellent auto and cross correlation properties of the code sequences.

### 3.2. LS Code Architecture

LS codes are defined as the combination of C and S subsequences, a Golay complementary pair, with zeros inserted to avoid overlapping between the two subsequences. If (C0, S0) and (C1, S1) are both Golay pairs of LS codes. As a result of inserted zeros, LS codes have features that aperiodic autocorrelation sidelobes and cross-correlations are zero within IFW zone [19, 20]. The main purpose of zeros insertion of the LS codes is to avoid the sequences C0 and C1 overlapping with the sequences S0 and S1. Note that it is also necessary to insert enough guard intervals between sequences with length longer than the maximum delay of the multipath channel [1]. The formation of proposed LS code presents in fig. 3.

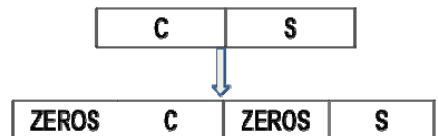

Fig. 3. Formation of LS Code

The LS code tree generated by the above steps in fig. 3 has the following properties.
(a) All the codes on the same layer of the code tree make up an LS code set;
(b) The auto correlation of each LS code is zero at non-zero shift.
(c) The two codes in the same node are completely complementary, i.e. their cross correlation is zero everywhere.
(d) If two nodes share the same father node, the IFW length of any two codes from the two nodes is equal to the sub-code length of the codes in the father node. Moreover, if two nodes share the same ancestor node, i.e. codes from the two nodes can be obtained through several recursive extensions from codes in the same ancestor node, the IFW length of any two codes from the two nodes is equal to the subcode length of the codes in the ancestor node [10]. The LS code tree structure is shown in fig. 4.

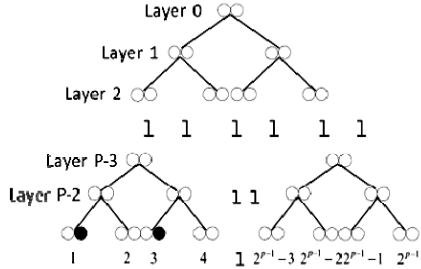

Fig. 4: Structure of the LS code tree

According to the above properties, the length of any two codes' IFW can be determined, if the code locations in the code tree are known. In a code tree, the nearer the two codes are, the longer their IFW will be.

### 3.3. Correlation properties of LS Code

The definitions and properties of auto and cross correlation functions (ACF/CCF) of any two codes LSi and LSj are given as:

$$........(1)$$

$$R_{ij}(\tau) = \sum_{n=0}^{N-1} C_{i,n} C_{j,(n+\tau) \bmod N} + \sum_{n=0}^{N-1} S_{i,n} S_{j,(n+\tau) \bmod N}$$

$$= \begin{cases} 2N, & \tau = 0, i = j \\ 0, & \tau = 0, i \neq j \end{cases}$$

Cross correlation properties of two LS code sequence has been observed in simulation result. It has also been observed that IFW width increased with code length. If the code lengths increase the IFW width has been increased. To determine the code length, the delay of multipath channel would be considered. The delay spread must be less then IFW to avoid any overlapping. simulation result of LS code correlation properties is shown in the result section.

### 3.4 Root raised cosine filter

Transmitting a signal at high modulation rate through a band-limited channel can create intersymbol interference. As the modulation rate increases, the signal's bandwidth increases. When the signal's bandwidth becomes larger than the channel bandwidth, the channel starts to introduce distortion to the signal. This distortion is usually seen as intersymbol interference. the spectrum of a rectangular pulse spans infinite frequency. In many data transmission applications, the transmitted signal must be restricted to a certain bandwidth. This can be due to either system design constraints. In such instances, the infinite bandwidth associated with a rectangular pulse is not acceptable. The bandwidth of the rectangular pulse can be limited, however, by forcing it to pass through a filter. The act of filtering the pulse causes its shape to change from purely rectangular to a smooth contour without sharp edges. Therefore, the act of filtering rectangular data pulses is often referred to as pulse shaping. Fig. 5 shows the Simulink design of RRC filter by using Xilinx System Generator blocks. System Generator is a system level modeling tool that facilitates FPGA hardware design and extends Simulink® in various ways in order to provide a powerful modeling environment that is well suited to hardware design. The tool provides high-level abstractions that are automatically compiled into a netlist code and also FPGA configuration bit stream file at the push of a button. System Generator block sets allow us to construct bit-accurate and cycle-accurate models of an FPGA circuit in Simulink.

In this proposed design, Root Raised Cosine (RRC) filter has been used for pulse shaping. Fig. 5 shows 32 order response of RRC filter. The specification of the RRC filter; sampling rate is 7.68 MHz, pulse duration 0.13 µs and the roll off factor is 0.25.

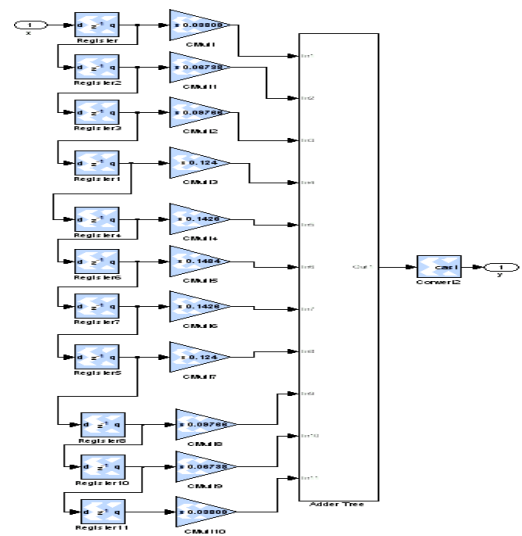

Fig.5. Simulink design of RRC filter.

### 3.5 Digital to analog converter

DAC converts the 16 bit digital data to analog. The output spectrum of DAC has shown in fig. 6.

The allowed BW for the transmission is around 16 MHz. DAC output image lies within the RF transmitter input BW range centered at 70 MHz.

The SAW filter in the RF transmitter will filter out the fundamental and all other images, except the image centered at 69.12 MHz shown in fig. 6. The RF Transmitter up-convert 70 MHz signal to 2.45 Ghz and transmit it through Antenna. Amplitude of the output spectrum reduces due to DAC sinc effect.

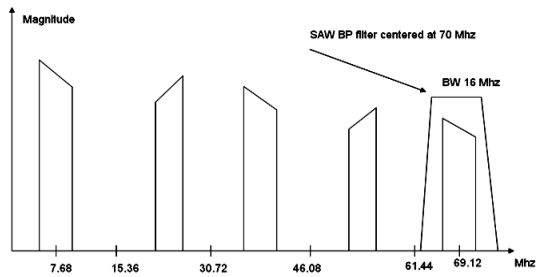

Fig. 6. DAC output spectrum.

## 4. HARDWARE IMPLEMENTATION

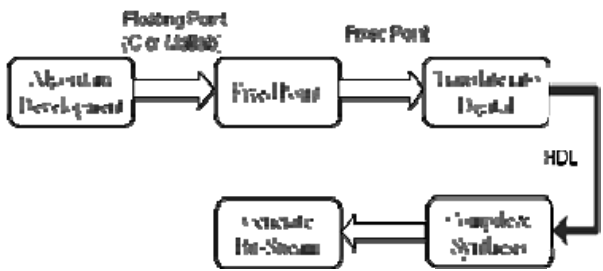

Fig. 7. Process flow diagram of hardware implementation

In the fig. 7, the total process flow of design and implementation has been illustrated. For LS Code generator design and to observe the correlation properties Matlab (ver: 7.a) is used. After simulation performed, expected auto and cross correlation properties of the proposed LS Code has been achieved. After LS Code generator design and investigate its correlation properties, LS Code sequence is converted from floating point to fixed point using Xilinx System Generator and AccelDSP. The remaining parts of the transmitter; up-sampler, RRC filter, Digital Up converter and Digital to analog converter is designed using Verilog. After HDL modeling the Verilog design has successfully compiled and synthesized by Xilinx ISE (ver: 9.2i). The Verilog code generates the NGC netlist using Xilinx ISE tools. The Xilinx system generator blocks in the Simulink environment also can be used to create HDL (Structural) netlist which can be translated to NGC netlist using Xilinx synthesis tool. In the both way Xilinx ISE tool generates the NGC netlist.

### 4.1 Experimental Setup

The simplified block diagram of BPSK transmitter shown in fig. 8. The transmitter is comprised of random number generator, up-sampler, Root Raised Cosine (RRC) filter, and digital up-converter (DUC) and digital to analog converter (16 bit DAC) and RF transmitter. The RF is configured to transmit at the center frequency of 2.45 GHz. The LS code generator is used as the data source for the transmitter. The RF board is based on Sundance SMT349 RF boards which consist of 2 separate Transceivers.

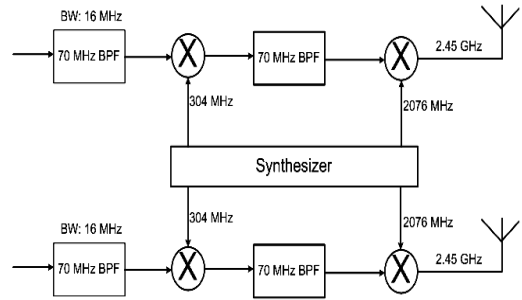

Fig. 8. Simplified block diagram of BPSK transmitter

Each Transceiver can be configured as either Transmitter or Receiver. Both transceivers configured as transmitter for this research. The RF transmitter up-convert 70 MHz signal to 2.45 Ghz and transmit it thru Antenna. The allowed BW for the transmission is around 16 MHz. The ADC is setup to run at 30.72 MHz so that the DAC output image lies within the RF transmitter input BW range centered at 70 MHz. The SAW band pass filter in the RF transmitter will filter out the fundamental and all other images, except the image centered at 69.12 MHz. Amplitude of the spectrum reduces due to DAC sinc effect. Fig.9 illustrates the connectivity between DAC and RF transceivers.

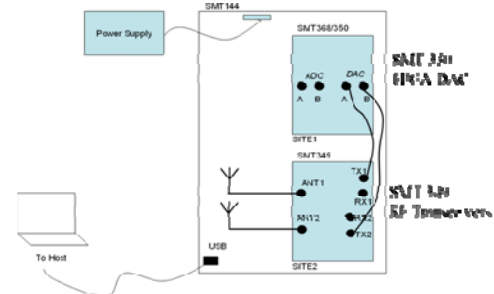

Fig.9. Process flow diagram of hardware implementation

LS code generator is designed and simulated by using Matlab (Matlab scripts included in appendix). Two sets of LS code is taken and converted in 16 bit format which are compatible with the hardware. The code sequence is combination of 1, -1 and 0. The code sequences converted as following:

TABLE 1
LS CODE SEQUENCE CONVERSION.

| Matlab Output | Converted form |
|---|---|
| 0 | 0.000000 |
| 1 | 1.000000 |
| -1 | -1.000000 |

After conversion of the LS code sequence it needs to be synthesized by using Xilinx ISE 9.2i. Both code sequences designed in Verilog by using lookup. After design the code sequences are compiled and synthesized in Verilog by using Xilinx ISE. After design and synthesized the both code sequence fed in to the FPGA by using 3L Diamond. The code set is used as sounding sequence for transmitter of channel sounder. Fig. 9 illustrates the connectivity between DAC and RF transceivers.

## 4.2 FPGA Task

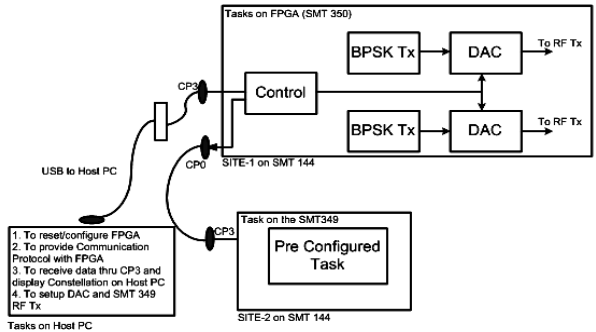

Fig.10. Tasks on FPGA

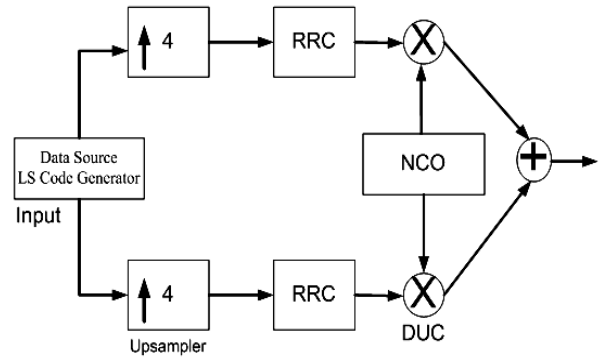

Fig.11. BPSK transmitter task

There are several task needs to be performed in FPGA shown in fig. 10 and each task is described briefly as follows:

**Control Task:** Control Task is responsible for reading the control data from Host and passes the control data to the ADC, DAC and clock synthesizers for proper operation of the data conversion. Control task also receives the control data for SMT349 which is then passed thru output port 0 on to the SMT349.

**DAC Task:** This task reads digital data from channel 0 (32 bit) and send data to DAC to converts the 16 bit digital data to analog. The DAC is comprised of the two channels DAC A and DAC B.

**BPSK Transmitter Task:** The BPSK transmitter task performed as shown in fig. 11. The source code is written in Verilog and synthesized by Xilinx ISE 9.2i is used to generate the configuration file (*.cfg file).

**Host Program:** The host program performs several tasks including resetting the hardware and configuring the FPGA on the SMT368, setting up the ADC, DAC, clock synthesizer and RF SMT349. The host program handles the required communication protocol to receive data (through USB) from FPGA and displaying the graphic data on the host PC.

**Steps to Download in FPGA from Host PC:** After compilation the host program is used for the followings:
- Reset and FPGA configuration
- ADC/DAC setup and clock synthesizer
- SMT349 RF transceivers setup
- To capture data stream sent from FPGA (via USB) to display constellation on Host PC.

## 5. RESULT AND DISCUSSION

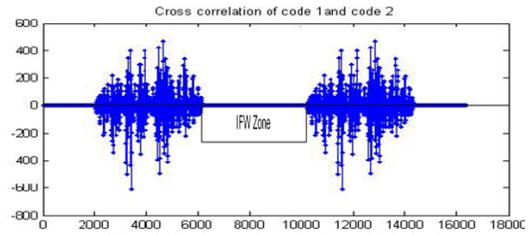

Fig.12(a). Cross Correlation of LS Code.

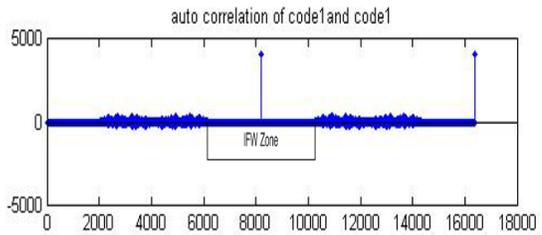

Fig. 12(b). Auto Correlation of LS Code.

Simulation result shows the auto correlation and cross correlation properties of LS code sequence length 8190 bits. Fig. 12(a) and fig. 12(b) presents the cross correlation and auto correlation result with 4000 bit length IFW of 8190 bit LS code sequence. With this code sequence its can measure the maximum delay of 4000 bit duration and minimum 1 bit duration.

Initially, MATLAB has been used, to design the LS code generator and investigate the correlation properties. After required codeset generates with excellent correlation properties, Xilinx system generator blocks used to design the Root Raised Cosine filter for pulse shaping purpose. The proposed design of transmitter has coded in Verilog, compiled and synthesized in ISE 9.2i. For Verilog modeling and implementation, the algorithm has to be thought of as a structural, behavioral, and physical version of the algorithm.

Fig. 13(a) shows the Register Transfer Level (RTL) schematic of the designed model. In integrated circuit design, RTL description is a way of describing the operation of a synchronous digital circuit. In RTL design, a circuit's behavior is defined in terms of the flow of signals (or transfer of data) between hardware registers, and the logical operations performed on those signals.

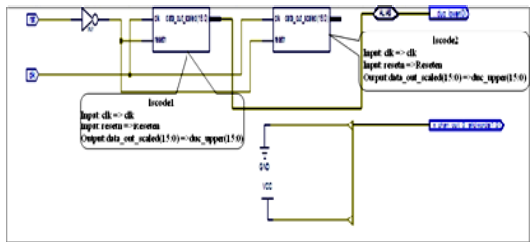

Fig.13 (a). RTL view of transmitter top module

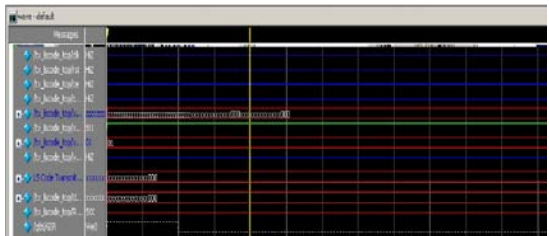

Fig.13(b). Simulation result of transmitter top module.

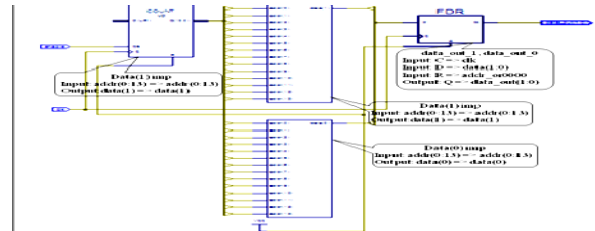

Fig.15(a). RTL view of Lscode2 Sub-module

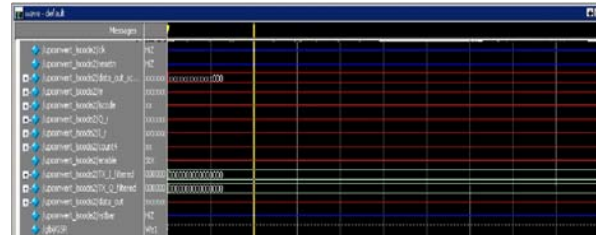

Fig.15(b). Simulation result of Lscode2 Sub-module

For the Verilog Simulation, the ISE 9.2i supports the ModelSim XE III 6.4b Waveform entity simulation methods' such as using Vector Waveform File (VWF). Fig. 13(a) presents the RTL schematic view of top module of proposed transmitter design. The transmitter top module consists of 2 sub-module lscode1 and lscode2. It has 2 input clk and resetn and 16 bit output data_out_scaled. Fig.13(b) shows the ModelSim simulation result of the transmitter top module.

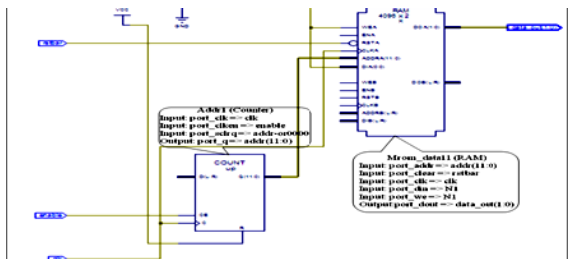

Fig. 14(a). RTL view of Lscode1 Sub-module.

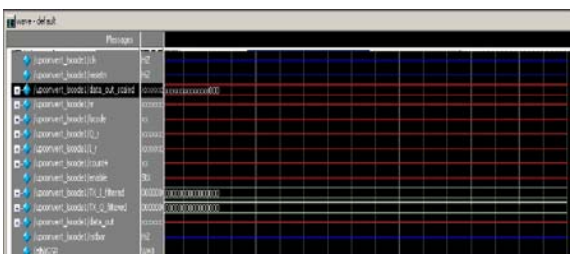

Fig. 14(b). Simulation result of Lscode1 Sub-module.

Fig.14(a) shows the RTL schematic view of *lscode1* sub-module. The *lscode1* sub-module consists of counter and memory. Counter has 3 inputs *clk*, clken and *port adr*_q; 12 bits output *port*_q. Fig. 14(b) shows the ModelSim simulation result of the lscode1 sub-module.

The RTL schematic view of lscode2 sub-module is shown in fig.15(a). The lscode2 sub-module consists of one 14-bit input and one bit output counter, two 14-bit input and one-bit output adder and 1 FDR. Counter has 3 inputs clk, clken and port adr_q; 12 bits output port_q. Fig.15(b) shows the ModelSim simulation result of the lscode2 sub-module.

## 6. APPLICATION TO RUN THE CONSTELLATION PROGRAM

Using Reset/Prog button in the constellation screen the FPGA has been configured and the status indicates the FPGA has been successfully configured. After successfully FPGA configured, RF setup has been done by using RF Setup button. Status shows "RF setup OK" if nothing goes wrong. ADC/DAC setup and clock synthesized by clicking the ADC/DAC SETUP button. Using ADCStream button data streaming captured from the BPSK output and display the constellation. Fig. 16 shows the constellation output of BPSK transmitter.

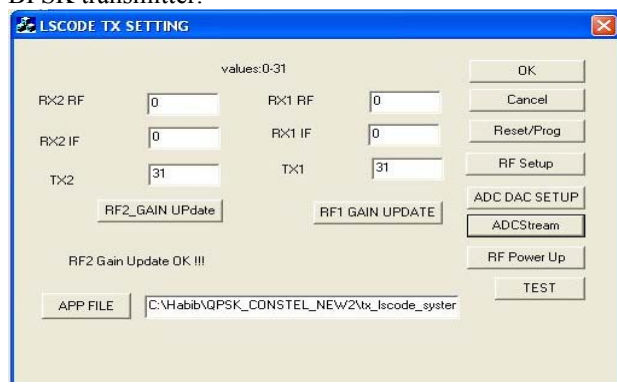

Fig.16. Constellation view of BPSK transmitter (TX1 & TX2)

## 7. SYSTEM PERFORMANCE

From design and implementation, it is studied that the IFWs of codes in a code set are not of the same length. In a CDM based system, all the active codes form a subset of the code set. Different subset demonstrates different IFW properties according to the length of the codeset.

In the above 2x2 transmitter design and implementation, two codeset has been chosen with same length. In matlab simulation, the length of the both LS Code sequence is 8190 has been observed and the length of IFW is 4000 bit.

Due to hardware limitations, the bandwidth of the system is 16MHz and chip rate is 7.68 MHz. Roll off factor of the RRC filter is 0.2 and the pulse duration of the RRC filter is 0.13µs (Pulse duration = 1/Chip Rate). The duration of the total IFW is 520 µs (total IFW Duration = Pulse duration * 4000). The proposed system can measure the delay of not smaller then 0.13 µs (1 bit duration) and not greater then 520 µs (total IFW duration).

## 8. CONCLUSION AND FUTURE WORK

This paper aims to design and implement LS code sequence generator that can be used as sounding sequence for CDM based 2 × 2 MIMO channel sounder. In this research, the correlation properties of proposed LS code have been investigated. Finally, the required both auto and cross correlation properties has been achieved and implemented in Xilinx FPGA. It is really challenging and important to achieve excellent auto and cross correlation properties simultaneously for CDM based channel sounding.

After design and implementation of transmitter using LS code this research can be extended to produce receiver to integrate in to a complete channel sounder. The performance of the system can be higher according to system bandwidth and code length.

**M. Habib Ullah** is M.Sc in Communication Engineering student in International Islamic University Malaysia. He is also a research assistant of Mobile Communication lab, tutor and demonstrator of Digital communication Lab at IIUM. He received B.Sc in Computer and Communication Engineering in 2004 from International Islamic University Chittagong, Bangladesh. He is the author several papers and his research interests include Wireless and Mobile Communications, CDMA, MIMO communication system, Interference Cancellation and Networking.

**Md. Niamul Bari** is M.Sc in Computer and Information Engineering student in International Islamic University Malaysia. He is also a Demonstrator of Networking lab and tutor of Electronics at IIUM. He received B.Sc in Computer Science and Engineering in 2004 from Asian University of Bangladesh. He is the author several papers and his research interests include Electronics, VLSI and RFID.

**A. Unggul Priantoro** obtained his B.Eng in electrical and computer engineering from Kobe University, Japan, in 1999 and his M.Eng and D.Eng degrees in Information Systems from Nara Institute of Science and Technology (NAIST), Japan, in 2001 and 2004 respectively. He was a post doctoral fellow at the same institute from April 2004 to Oct. 2004 under Center of Excellence (COE) program. He joined the Electrical and Computer Engineering Dept., IIUM in 2004. He served as Deputy Director of the IT Division from June 2006 until May 2008. His research interests are in wireless communications and networking.